\documentclass[aps,prb,showpacs,preprint,amsmath,amssymb,superscriptaddress]{revtex4}
\usepackage{amsmath,amssymb,epsfig,color,textcase}

\begin{document}

\title[First principles investigation of exchange interactions]{First principles investigation of exchange interactions
in quasi-one-dimensional antiferromagnet CaV$_2$O$_4$}

\author{Z.V.~Pchelkina}
\email{pzv@ifmlrs.uran.ru}
\affiliation{Institute of Metal Physics, S.Kovalevskoy St. 18, 620990 Ekaterinburg, Russia}
\affiliation{Ural Federal University, Mira St. 19, 620002 Ekaterinburg, Russia}

\author{ I. V. Solovyev }
\email{SOLOVYEV.Igor@nims.go.jp}
\affiliation{Ural Federal University, Mira St. 19, 620002 Ekaterinburg, Russia}
\affiliation{Computational Materials Science Unit, National Institute for Materials Science, 1-1 Namiki, Tsukuba, Ibaraki 305-0044, Japan}

\date{\today}

\begin{abstract}
The effect of orbital degrees of freedom on the exchange interactions in the spin-1 quasi-one-dimensional antiferromagnet
CaV$_2$O$_4$ is systematically studied. For this purpose a realistic low-energy model with the parameters derived from the
first-principles calculations is constructed.
The exchange interactions are calculated using both the theory of infinitesimal spin rotations near the mean-field
ground state and the superexchange model, which provide a consistent description. The obtained behavior of exchange
interactions substantially differs from the previously proposed phenomenological picture based on the magnetic measurements
and structural considerations, namely: (i) Despite quasi-one-dimensional character of the crystal structure, consisting of the
zigzag chains of
edge-sharing VO$_6$ octahedra, the electronic structure is essentially three-dimensional, that leads to finite interactions
between the chains; (ii) The exchange interactions along the legs of the chains appear to dominate; and (iii) There is a
substantial difference of exchange interactions in two crystallographically inequivalent chains. The combination of these
three factors successfully reproduces the behavior of experimental magnetic susceptibility.

%\noindent{\it Keywords}: CaV$_2$O$_4$; exchange interaction; electronic structure; low-energy model; low-dimensional magnetism; magnetic frustration

\end{abstract}

%\submitto{\JPCM}

\maketitle

\section{Introduction \label{intro}}

\begin{figure}[!]
\resizebox{16cm}{!}{\includegraphics{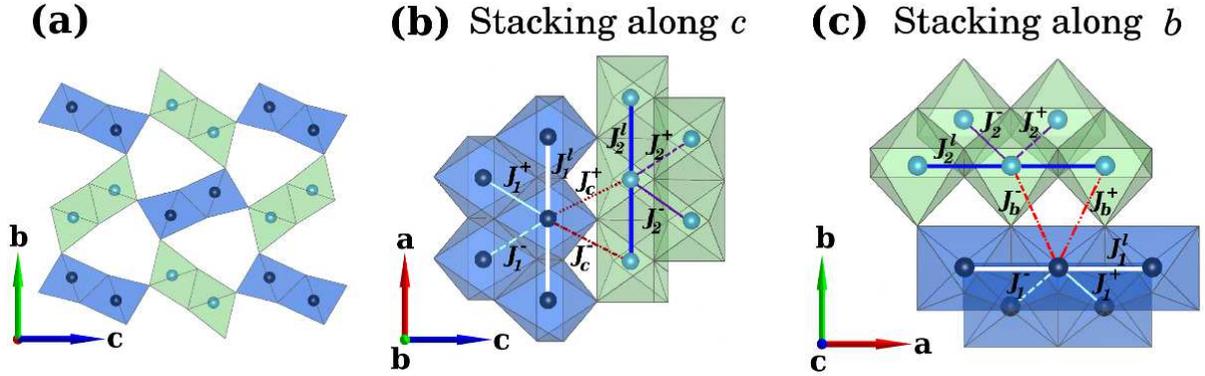}}
\caption{\label{cavo_str}(Color online)
The crystal structure of monoclinic phase of CaV$_2$O$_4$ in \emph{bc}, \emph{ac} and \emph{ba} projections.
The two nonequivalent V atoms are shown as black (V1) and cyan (V2) spheres. The oxygen atoms in the corners of octahedra and
the Ca atoms are not shown for simplicity. The oxygen octahedra around V1 are colored in blue, while octahedra around V2 are colored in light green.
Each zigzag chain
of vanadium atoms has two neighboring chains of other type, stacking along
monoclinic directions $c$ and $b$, as explained in panels (b) and (c), respectively.
The definition of main exchange interactions for each stacking is shown.
For the visualization, the VESTA software~\cite{MommaK.Izumi2011} was used.}
\end{figure}

  The CaV$_2$O$_4$ compound was studied both theoretically and experimentally owing to
its low-dimensional magnetism and frustrated structure~\cite{Kikuchi2001, Niazi2009, Pieper2009}. The high temperature orthorhombic phase
(the space group $Pnam$)
undergoes the phase transition into low temperature monoclinic phase
(the space group $P2_1/n11$) at T$_s \approx 141$ K.
The main motif of both structures is zigzag double chains of edge-sharing VO$_6$ octahedra (see figure~\ref{cavo_str}).
The distances between nearest and next-nearest V neighbors are nearly equal, which together with the
antiferromagnetic (AFM) type of interactions gives rise to the geometrical frustration.
The electronic configuration of V$^{3+}$ is 3$d^2$, making CaV$_2$O$_4$
the appropriate compound for investigation of S=1 quasi-one-dimensional magnetism.

   CaV$_2$O$_4$ has two crystallographically inequivalent types of vanadium atoms, V1 and V2, forming the zigzag chains.
The vanadium atoms are displaced out of the center of octahedra (figure~1 of supplemetary material avaliable at~\cite{SM}),
yielding the existence of finite electric dipoles. However, both structures possess the inversion symmetry,
meaning that the dipoles are ordered antiferroelectrically.
Each zigzag chain propagates along the $a$ axis and
has two neighboring chains of other type, which are stacked along $c$ and $b$ axes
(see figures~\ref{cavo_str}(b) and \ref{cavo_str}(c), respectively). Each vanadium atom has six nearest vanadium neighbors forming
six main exchange paths (the notations of corresponding interatomic distances are given in the brackets):
$J^l_1$ and $J^l_2$ ($d^l_1$ and $d^l_2$) - along the ``leg'' of the chain formed by V1 and V2, respectively; $J^+_1$, $J^-_1$ and $J^+_2$, $J^-_2$
($d^+_1$, $d^-_1$ and $d^+_2$, $d^-_2$) - along the zigzag in the positive and negative direction of $a$
(denoted by ``$+$'' and ``$-$'', respectively); and two groups of interchain interactions
along $c$ and $b$: ($J^+_c$, $J^-_c$) and ($J^+_b$, $J^-_b$), respectively
(see figure~\ref{cavo_str}).

  The magnetic structure of CaV$_2$O$_4$ have been studied already in 70's~\cite{Bertaut1967, Hastings1967}
but it was impossible in that time to resolve
the low temperature monoclinic crystal structure and analyze correctly experimental data. It is well known that below
T$_N\approx$51-78 K the long range antiferromagnetic order with a propagation vector $\mathbf {k}=(0,\frac{1}{2}, \frac{1}{2})$
sets in~\cite{ Niazi2009, Bertaut1967, Hastings1967, Sugiyama2008, Zong2008}. The reduction of magnetic moment
$1.0  \mu_B \leq  m  \leq  1.59  \mu_B$ was detected in $^{51}$V nuclear
magnetic resonance (NMR) measurements, muon-spin spectroscopy investigations and powder
diffraction~\cite{Niazi2009, Bertaut1967, Hastings1967, Sugiyama2008, Zong2008}.
The collinear spin orientation was found in the first neutron powder diffraction experiments at low temperatures~\cite{Bertaut1967, Hastings1967}.
However, it was questioned in more recent NMR measurements~\cite{Zong2008} and neutron diffraction experiments on high quality
single crystals~\cite{Pieper2009}, which suggest some noncollinear spin arrangement.

  Since the distances in the leg $d_{1,2}^{l}$ are nearly the same as
in the zigzag $d_{1,2}^{\pm}$ (see figure~\ref{cavo_str}) one could naively expect nearly equal exchange interactions
$J_{1,2}^l \approx J^{\pm}_{1,2}$. However, the high-temperature dc susceptibility
measurements of CaV$_2$O$_4$ single crystals reveal that above T$_s$ the system behaves like a S=1 Heisenberg chain~\cite{Pieper2009}.
In order to explain this fact, it was typically assumed that (i) The zigzag chains, formed by V1 and V2, are nearly equivalent; and
(ii) Three $t_{2g}$ orbitals of vanadium sites are oriented in such a way that their lobes are
parallel to $d^l$, $d^+$, and $d^-$, giving rise to the exchange paths $J^l$, $J^+$, and $J^-$, respectively
($J_{leg}$, $J'_{zz}$ and $J''_{zz}$ in the notations of~\cite{Pieper2009}).

  In the orthorhombic phase, the lowest $t_{2g}$ orbital was supposed to be occupied by one electron, while the second electron
resides on a double degenerate level~\cite{Pieper2009}. The direct overlap of the first orbitals
leads to the strong AFM interaction along the leg ($J^l$), while the interactions in the zigzag are
identical ($J^+$=$J^-$) and should be considerably weaker than $J^l$. Hence, the system in the orthorhombic phase
could be considered as a S=1 Haldane chain. This scenario was supported by results of exact diagonalization calculations,
supplemented with the fitting of the experimental high-temperature susceptibility data,
which yield $J^l = -$$18.60$ meV and $J^+ = J^- = -$$3.02$ meV~\cite{Pieper2009}. Nevertheless, in the same work~\cite{Pieper2009},
yet another scenario was proposed with $J^+ = J^- = -$$19.85$ meV and $J^l = -$$0.75$ meV, which equally well fits the
experimental data. In both cases, the fitting yields the Curie temperature $\Theta =-$$418$ K and the
effective magnetic moment $\mu_{exp} =2.77 \mu_B$~\cite{Pieper2009}.

  Under the transition to the monoclinic phase, the additional distortions of the VO$_6$ octahedra completely
lifts the degeneracy of $t_{2g}$ orbitals and make all the exchange paths inequivalent. Moreover, two
$t_{2g}$ orbitals are supposed to be occupied and one orbital is empty, that should lead to the inequality $(J^l,J^+) \gg J^-$.
Such magnetic structure with the strong exchange couplings along the leg and every second interaction along the zigzag
corresponds to the spin-1 ladder.

  The above scenario was based solely on the qualitative structural consideration without taking into account the
existence of two nonequivalent
types of vanadium atoms. Moreover, the assumed type of the orbital ordering was purely empirical and had no
proper link to details of the crystal structure.
In the present work we report theoretical investigation of electronic structure, orbital ordering and
exchange interactions in CaV$_2$O$_4$.
For this purpose we construct the realistic low-energy model and derive parameters of
this model from the first-principles electronic structure calculations. Then we solve the model and obtain
parameters of interatomic exchange interactions. The calculated values of exchange integrals are analyzed within superexchange theory.
The spin model with
the calculated exchange parameters was solved by quantum Monte-Carlo method in order to compare theoretical results
with the experimental magnetic susceptibility.

\section{Method}
\begin{figure}[!htb]
\centerline { \resizebox{8cm}{!}{\includegraphics[angle=270]{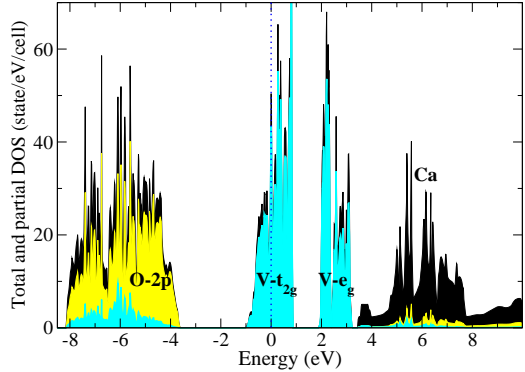}}}
 \caption{\label{cavo_dos}(Color online)
LDA total and partial densities of states for the monoclinic phase of CaV$_2$O$_4$. The black area stands for the total DOS,
yellow - for O-2$p$ and cyan - for V-3$d$ partial DOS. The Fermi level (dotted line) corresponds to zero.}
\end{figure}

In order to analyze the electronic and magnetic properties of CaV$_2$O$_4$ we employ the previously developed method of ``realistic
modeling'' (see~\cite{downfolding} for a review).
The same approach has been performed for the theoretical investigation of related quasi-one-dimensional compound NaV$_2$O$_4$~\cite{navo_prb}.
First, the band structure of CaV$_2$O$_4$ was calculated in the local density approximation (LDA). The total and partial densities of states (DOS)
for the monoclinic phase are shown in figure~\ref{cavo_dos}. The bands
located near the Fermi level have V-$t_{2g}$ character.
Therefore, we consider the behavior of only these low-energy bands and construct for them Hubbard-type model.
All parameters of this model can be derived from the
first-principles electronic structure calculations in the Wannier basis. All computational details can be found in~\cite{downfolding}.

Most of calculations reported in this work are performed for the experimental monoclinic $P2_1/n11$
structure (unless it is specified otherwise).
We use the data
from~\cite{pieper_diploma}, but transform them to the conventional setting with
the unique axis $a$ and monoclinic angle $\beta$. The corresponding lattice parameters are
$a =2.99780$~\AA, $b =9.19524$~\AA, $c =10.68025$~\AA~and $\beta =90.767^{\circ}$, and all atomic coordinates
are summarized in supplementary material avaliable at~\cite{SM}.

To construct the model one needs to specify the three sets of parameters, namely, the crystal field (CF),
transfer integrals, and screened Coulomb interactions.
\begin{figure}[!htp]
\centerline { \resizebox{7cm}{!}{\includegraphics[angle=270]{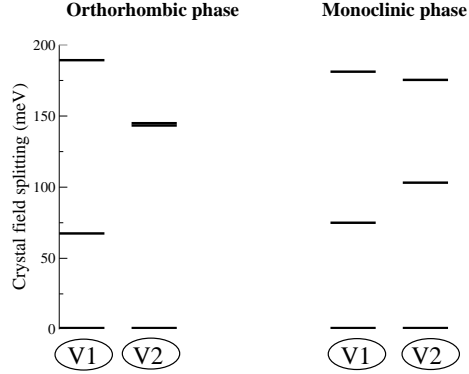}}}
 \caption{\label{cavo_cf}
The crystal field splitting of three $t_{2g}$ states (in meV) for two types of V atoms in
orthorhombic (left) and monoclinic (right) structures of CaV$_2$O$_4$.}
\end{figure}
The CF splitting of the three $t_{2g}$ levels for the orthorhombic and monoclinic structures is shown in figure~\ref{cavo_cf}.
The relative position of atomic $t_{2g}$ levels in the orthorhombic phase is (0, 67, 189) meV and (0, 143, 144) meV,
while in
the monoclinic phase it is (0, 75, 181) meV for V1 and (0, 103, 175) meV for V1 and V2, respectively.
Thus, in
the orthorhombic case two of the three levels of the V2 ion are almost degenerate while in the monoclinic
phase this degeneracy is lifted by the additional distortion.

\begin{figure}[h!]
\begin{center}
\includegraphics[width=4.2cm]{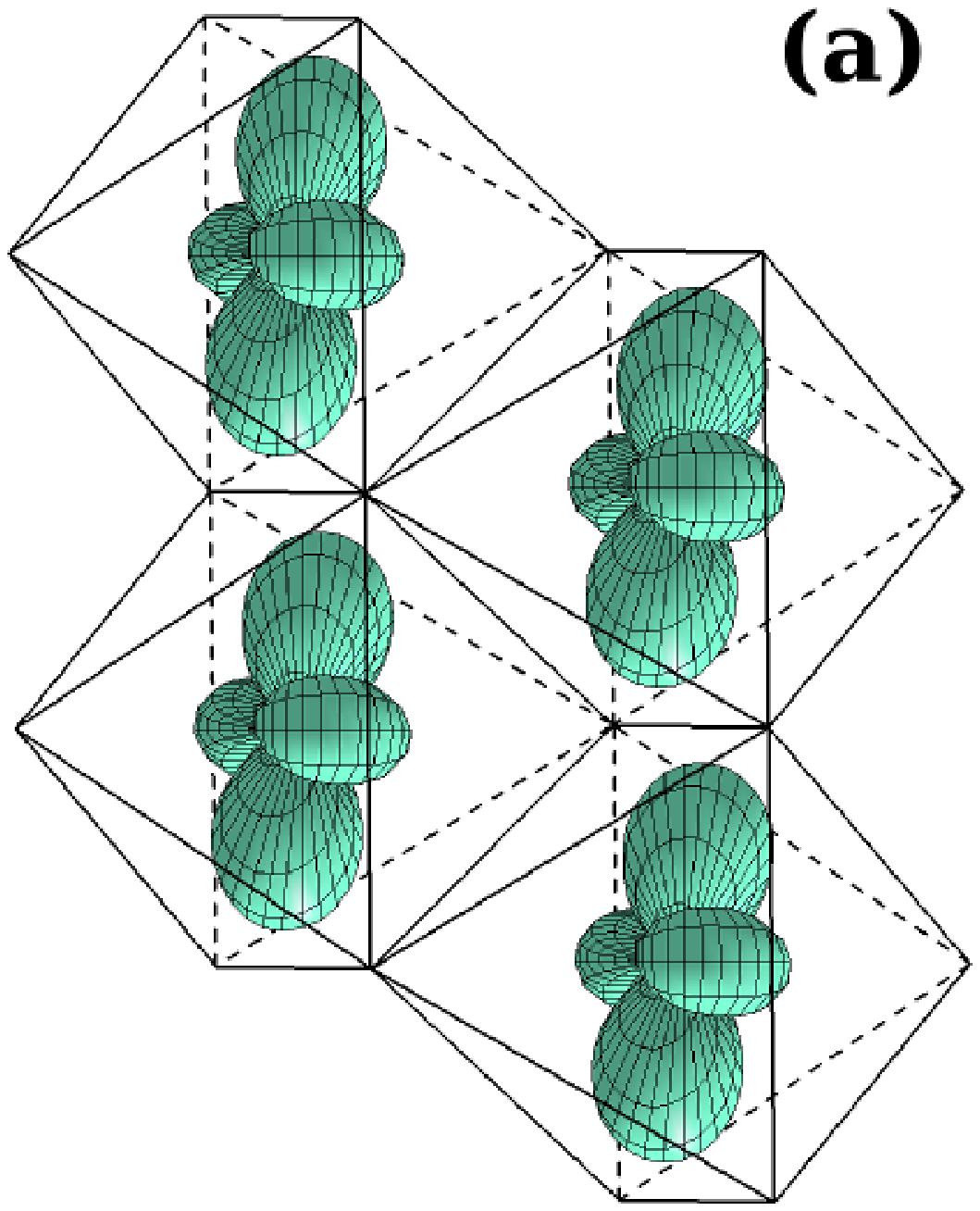}
\includegraphics[width=4.2cm]{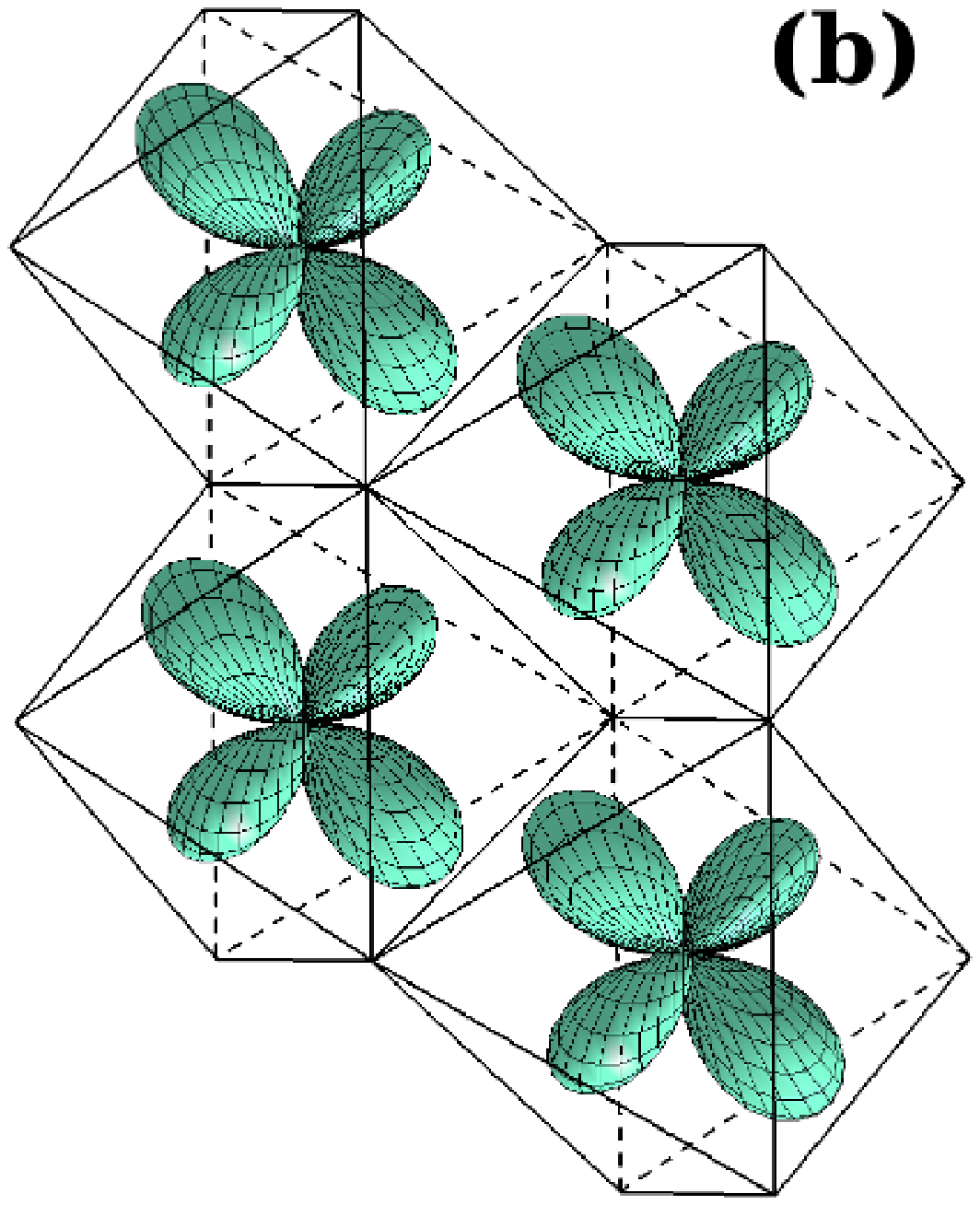}
\includegraphics[width=4.2cm]{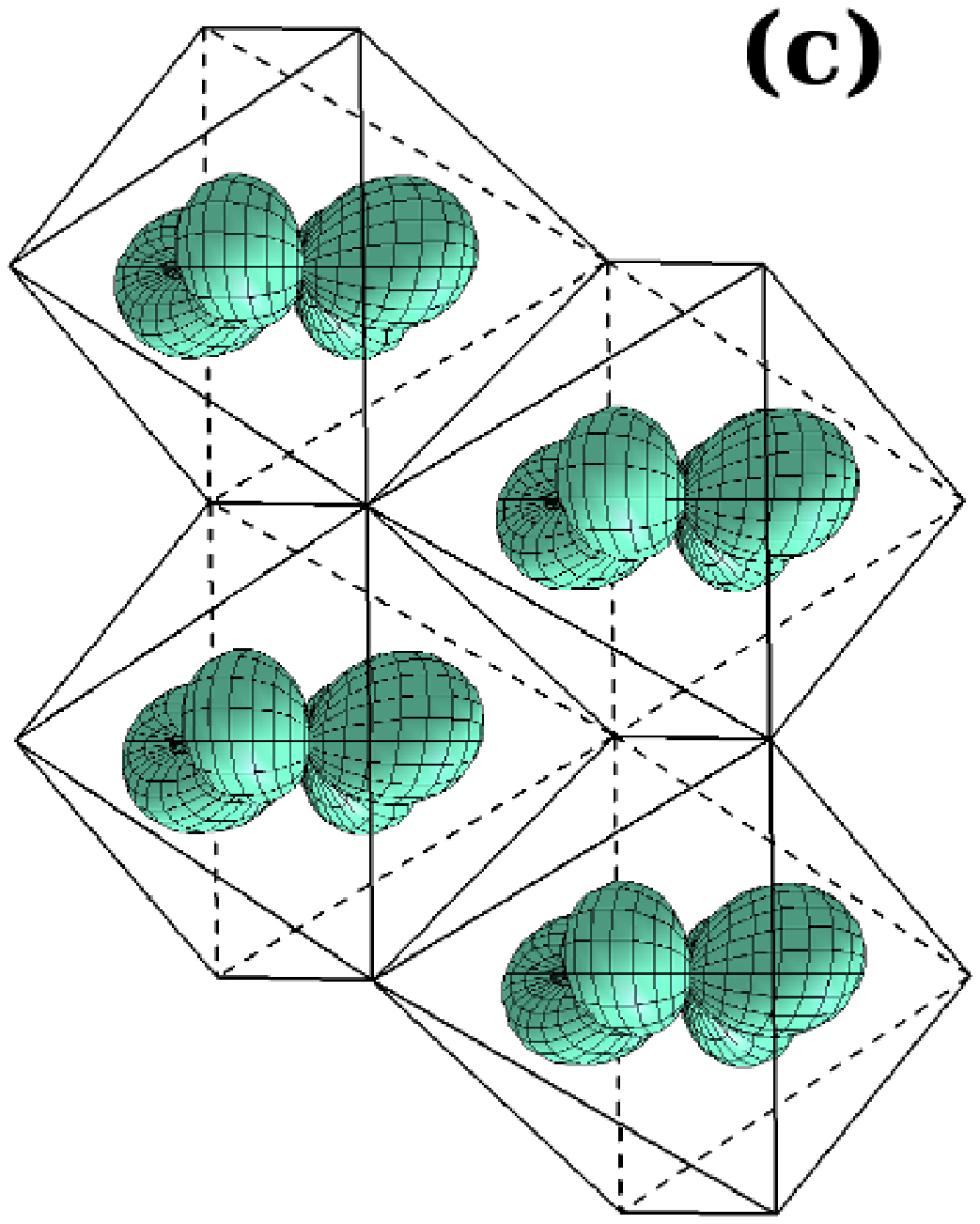}

\includegraphics[width=4.2cm]{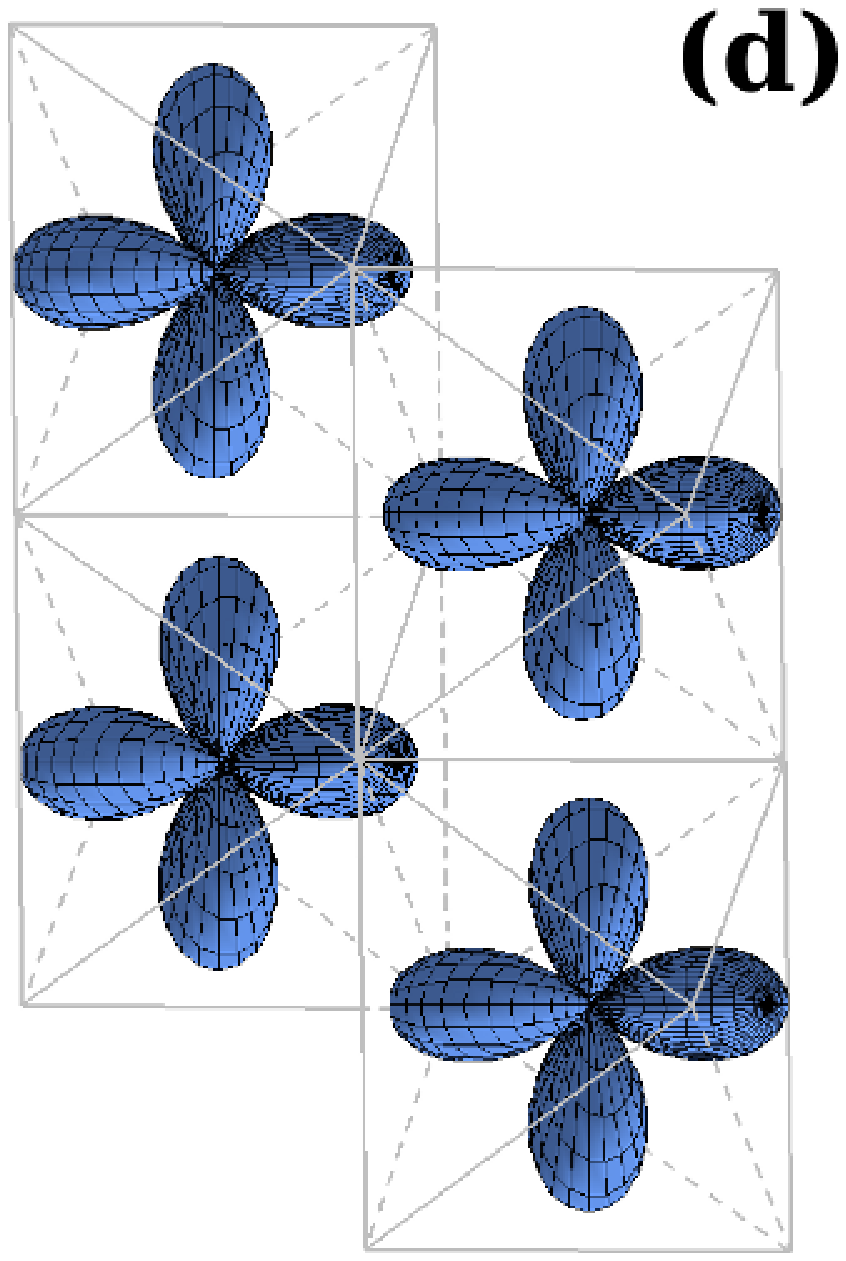}
\includegraphics[width=4.2cm]{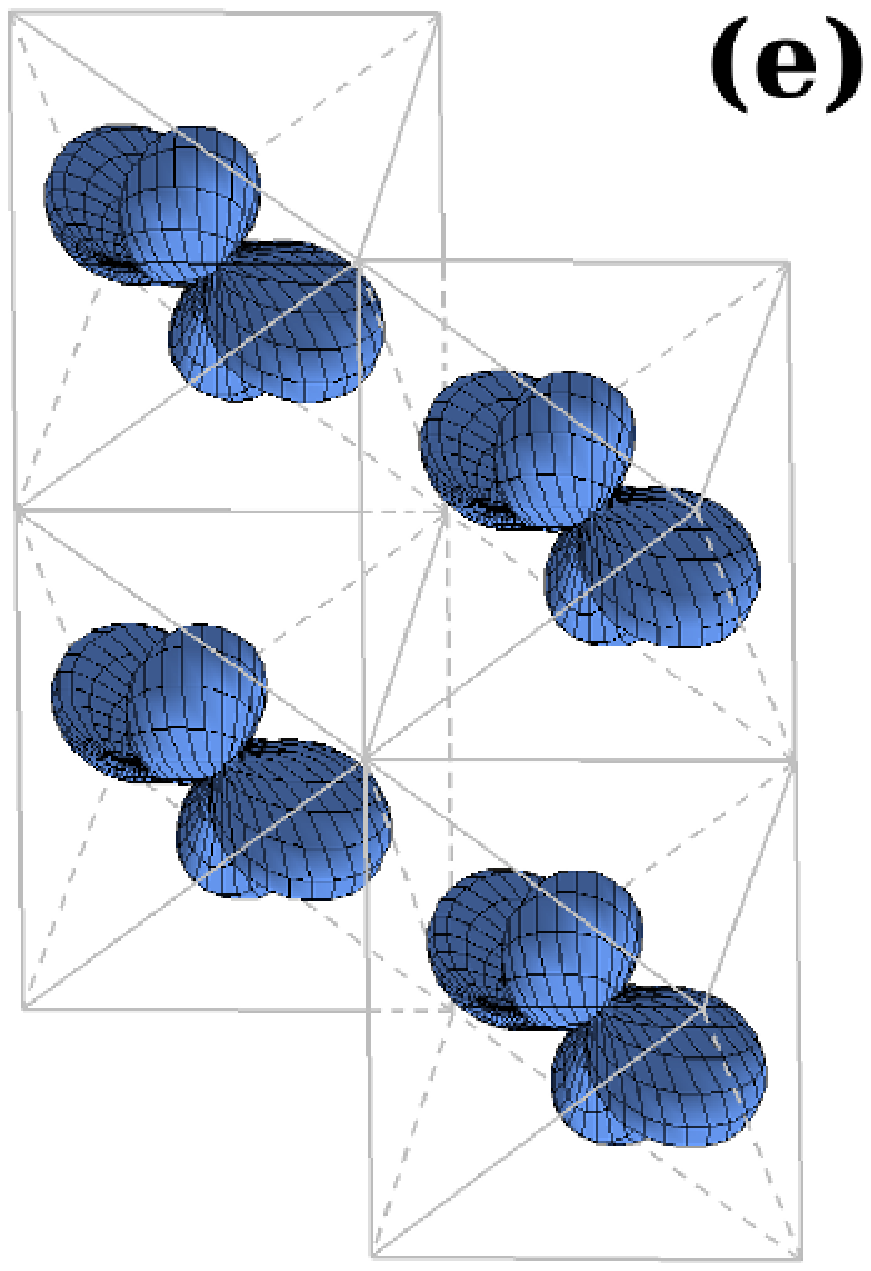}
\includegraphics[width=4.2cm]{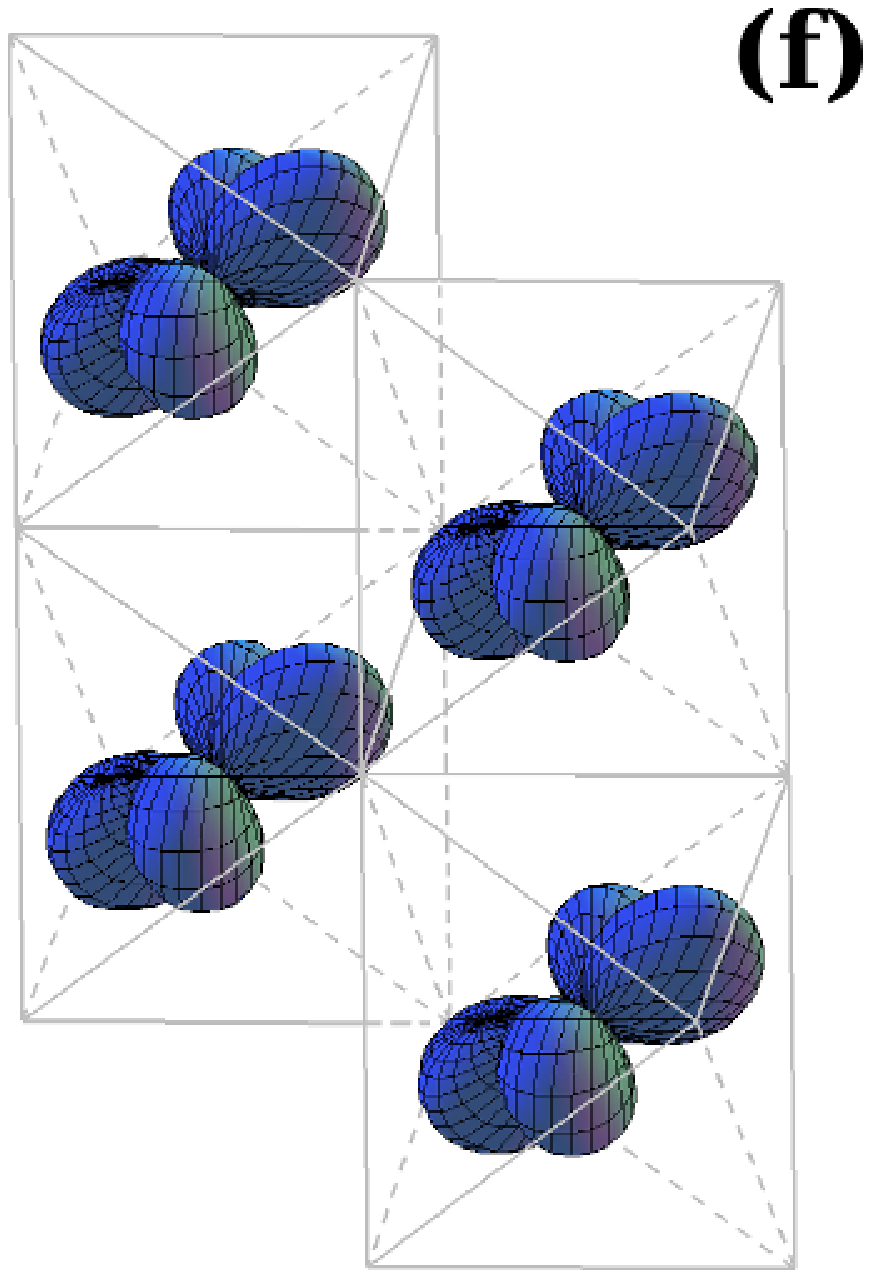}
\end{center}
\caption{\label{mono_orb} (color online).
The three $t_{2g}$ orbitals corresponding to the crystal field levels shown in figure~\ref{cavo_cf} for monoclinic phase of CaV$_2$O$_4$.
The orbitals (a), (b), (c) correspond to V1, (d), (e), (f) correspond to V2. The orbitals are shown in order of increasing their energy
from left (the lowest in energy orbital) to right (the highest in energy orbital).}
\end{figure}

The arrangement of these three $t_{2g}$ orbitals in monoclinic phase of CaV$_2$O$_4$ corresponding to
the aforementioned crystal field levels is illustrated in figure~\ref{mono_orb} in global coordinate frame.
In the following, for each vanadium site $i$ (which can be either V1 or V2) we will denote the lowest, middle, and highest
$t_{2g}$ orbitals as $\phi_i^1$, $\phi_i^2$, and $\phi_i^3$, respectively.

  The magnetic interactions are intimately connected with the spacial ordering of the  $t_{2g}$ orbitals~\cite{Kugel1982}.
Small structural distortions can lead to the significant changes
in the orbital ordering and magnetic properties of such compounds.
In the monoclinic CaV$_2$O$_4$, there are two $d$ electrons occupying the two lowest orbitals $\phi^1_i$ and $\phi^2_i$.
By neglecting for a while small off-centering of the vanadium ions, all VO$_6$ octahedra are compressed along
the shortest V-O-V distance, which can be denoted as local $z$ axis and, for V1, almost coincides with the crystallographic $c$ axis.
Then, the orbital with $xy$ symmetry should be the lowest in energy.
Indeed all $\phi^1_i$ orbitals have predominantly $xy$ character in agreement with these simple structural consideration (see figure~\ref{mono_orb}).
The lobes of the $\phi^1_i$ orbitals on the neighboring V ions in the $a$ direction are pointed along the leg of the zigzag chain.
Hence, one could expect large transfer integrals in the legs of the zigzag chains.

  The $3$$\times$$3$ matrices of transfer integrals
$t^{mm'}_{ij}$, calculated
in the local CF representation are summarized in table III and IV of supplementary material avaliable at~\cite{SM}.
In these notations, $i$ and $j$ denote the vanadium sites, which can be of the type V1 or V2,
while $m$ runs over the CF orbitals $\phi_i^1$, $\phi_i^2$, and $\phi_i^3$. As expected, the largest transfer integrals
operate between $\phi^1$ orbitals in the legs of the zigzag chains ($t_{ij}^{11} = -$$265$ and $-$$233$ meV for the
chains formed by V1 and V2, respectively).
The transfer integrals between the chains are weaker, but comparable with the intrachain ones.
Thus, despite
the quasi-one-dimensional character of the crystal structure, the transfer integrals
in CaV$_2$O$_4$ are essentially three dimensional.
The same trend have been found for related quasi-one-dimensional compound NaV$_2$O$_4$~\cite{navo_prb}.

 In order to compute the screened Coulomb interactions in the $t_{2g}$ band we use the following procedure~\cite{downfolding}. First we apply
constrained LDA to take into account the screening of atomic orbitals. Then the random-phase approximation (RPA) was employed
to take into account the self-screening by the same 3$d$-electrons which participate in the formation of other bands due to the hybridization effects.
The fitting of screened interactions in terms of two Kanamori parameters~\cite{Kanamori1963} results in the following values of the
intraorbital Coulomb interaction $U$=3.42 (3.46) eV and the intraatomic exchange coupling $J_H$=0.63 (0.64) eV for V1 (V2).

\section{Results and discussion}
\subsection{Exchange interactions and magnetic ground state}

\begin{figure}[h!]
\begin{center}
\includegraphics[width=8cm]{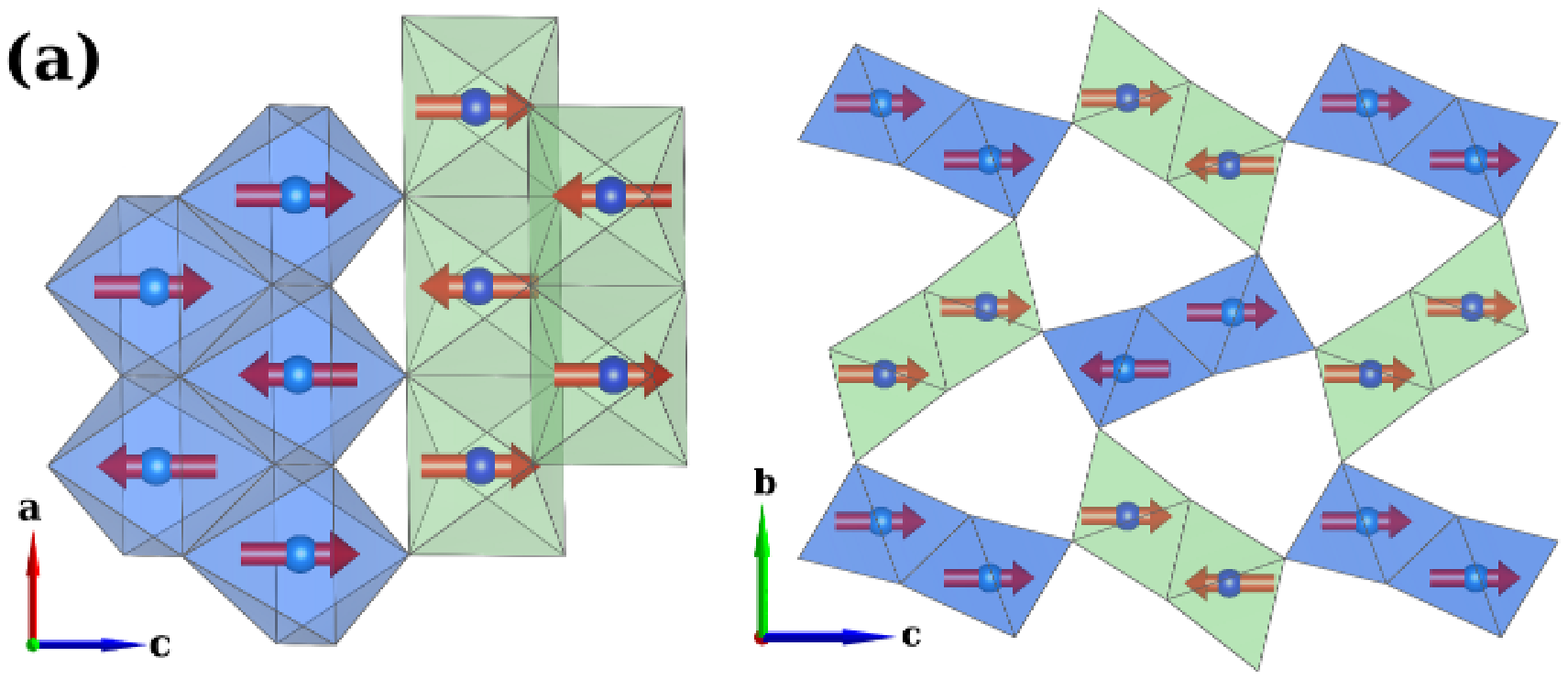}
\includegraphics[width=8cm]{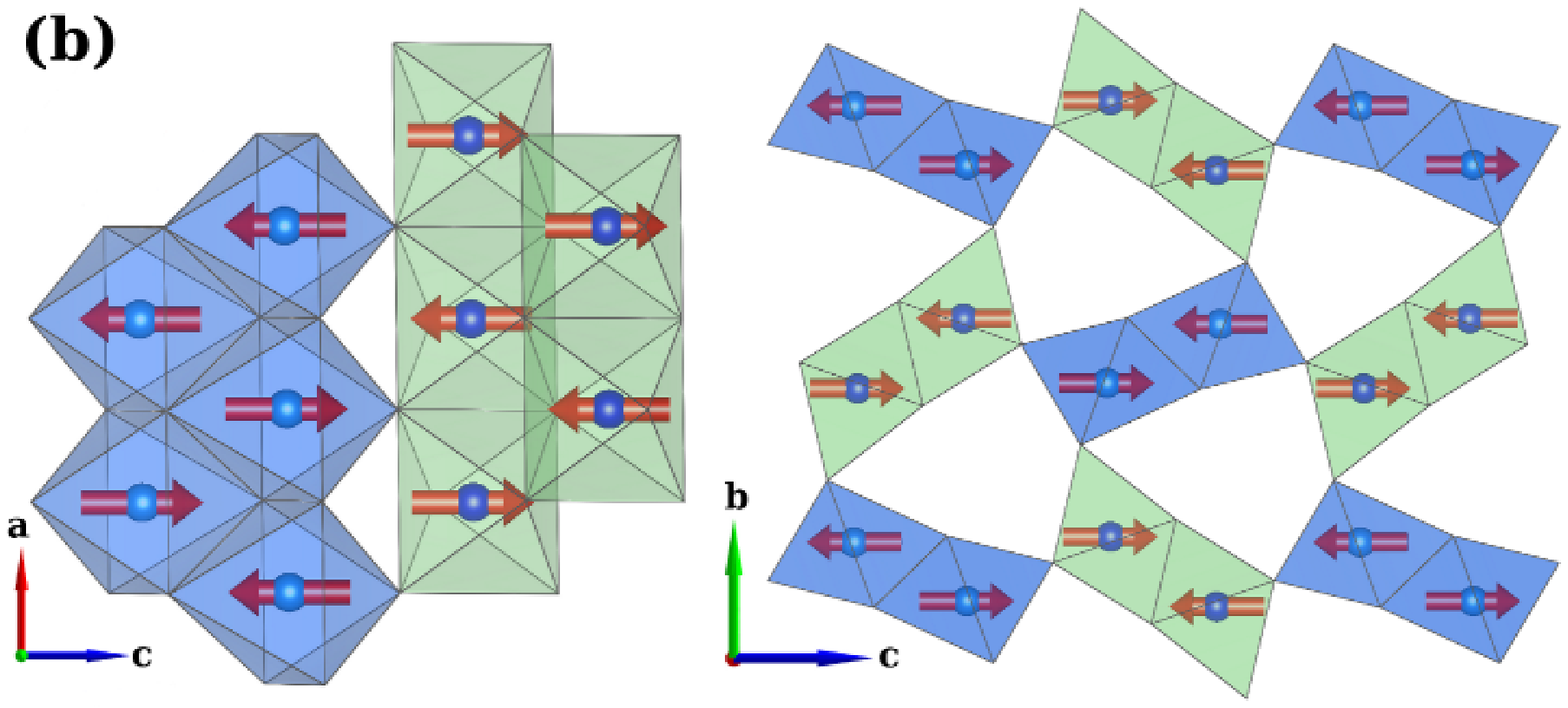}
\includegraphics[width=8cm]{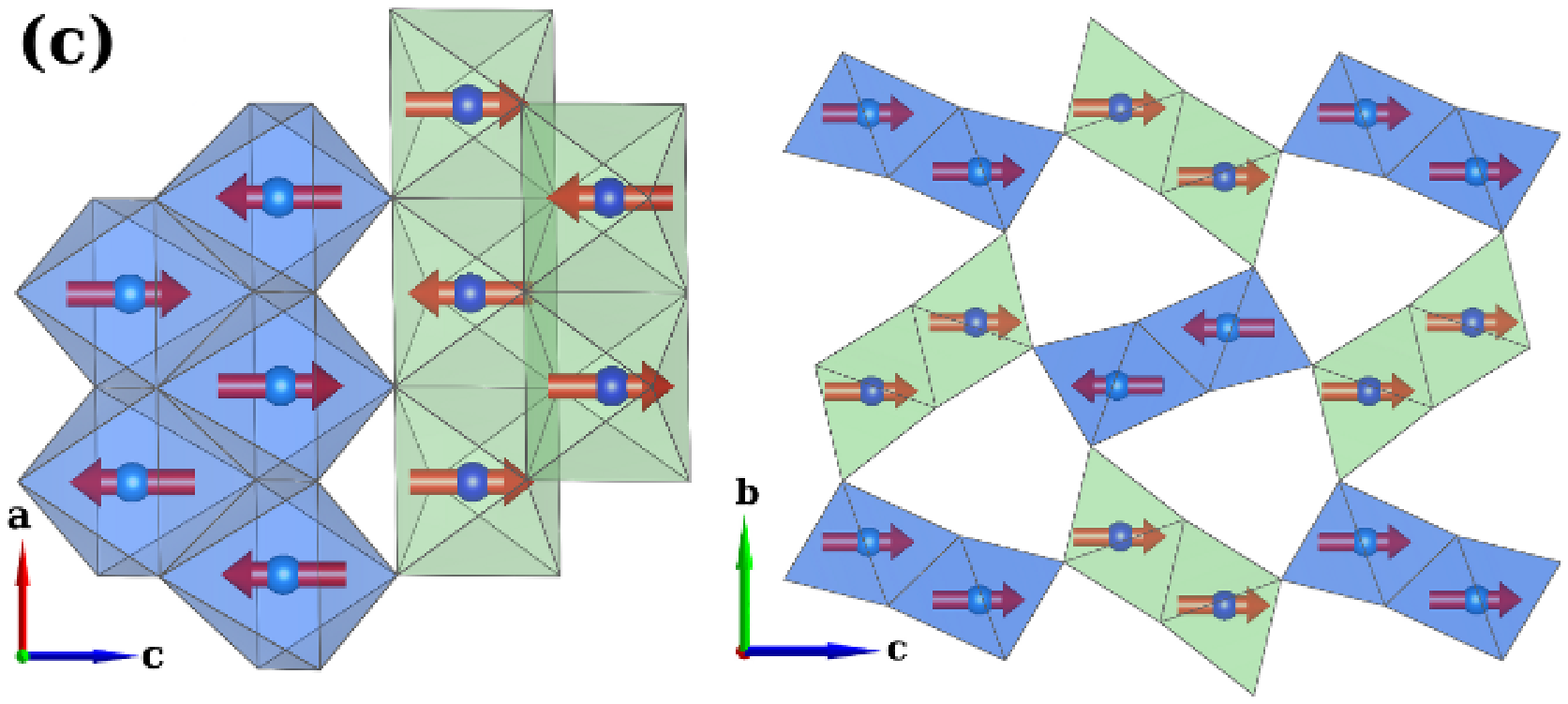}
\end{center}
\caption{\label{afm_double} (Color online).
The sketch of different AFM configurations for monoclinic phase of CaV$_2$O$_4$ in the cell, doubled along $a$ axis: AFM1 (a), AFM2 (b) and AFM3 (c). }
\end{figure}

  First, we solve the obtained low-energy electron model in the mean-field Hartree-Fock approximation.
For these purpose, we consider four collinear magnetic configurations, two of which, AFM2 and AFM3, was reported to be in moderate agreement with
the single crystal neutron diffraction data (see figures 5.29(a) and 5.29(b) in ~\cite{Pieper2009}). The unit cell was doubled
along the $a$ axis in order to arrange the V spin moments antiferromagnetically as was
detected in the single-crystal neutron diffraction experiments~\cite{Pieper2009, pieper_diploma}.
The sketch of the three considered AFM arrangements is shown in figure~\ref{afm_double}.

The parameters of interatomic magnetic interactions were calculated for different magnetic configurations
by applying the perturbation theory expansion with respect to the infinitesimal spin rotations near the equilibrium state~\cite{Liechtenstein}.
This procedure corresponds to the \textit{local} mapping of the total energy change associated with the small
rotations of spins onto the Heisenberg model
\begin{equation}
\label{spin_model}
H=-\sum_{i>j}J_{ij}{\mathbf e}_i {\mathbf e}_j,
\end{equation}
where ${\bf e}_i$ is the direction of spin at the site $i$.
The results of exchange interaction calculation for monoclinic CaV$_2$O$_4$ in
the ferromagnetic (FM) and three
AFM configurations are summarized in table~V of supplementary material avaliable at~\cite{SM}.
Since the degeneracy of $t_{2g}$ orbitals is lifted by the lattice distortion, these
exchange integrals only weakly depend on the type of the magnetic order in which they are calculated,
that justifies the use of the spin-only model.

The leading exchange interactions $J^l_1 =-19.9$ meV, $J^l_2 =-13.9$ meV correspond to the strong AFM coupling along the leg
of the zigzag chains and are about 7 times larger than the
remaining interactions. The parameters of antiferromagnetic interactions in the zigzag-rung
are $J^+_1 =-1.3$ meV, $J^-_1 =-0.4$ meV and $J^+_2 =-1.6$ meV, $J^-_2 =-1.6$ meV for V1 and V2, respectively. Such a behavior
corresponds to the limit $J^l \gg J^{\pm}$, which is consistent with the analysis of experimental magnetic susceptibility
data in~\cite{Pieper2009}.
Nevertheless, the interactions $J^l_1$ and $J^l_2$ in two different types of chains are substantially different.

  The interactions between different types of the zigzag chains along $c$ direction are ferromagnetic: $J^+_c =1.1$ meV and $J^-_c =2.9$ meV.
Similar interactions along $b$ are found to alternate: $J^+_b =-1.3$ meV is antiferromagnetic, while $J^-_b =1.5$ meV
is ferromagnetic. This behavior
should correspond to the AFM3 magnetic alignment. This result is totally consistent with direct Hartree-Fock calculations,
where the AFM3 state was found to have the lowest energy.

  In order to get some insight into microscopic origin of exchange interactions, one can also estimate the parameters
in the superexchange approximation, starting from the atomic limit and considering virtual hoppings to the
neighboring sites in the first order of $1/U$~\cite{Kugel1982, Anderson1959}.
Then, $J_{ij}$ can be calculated as the energy difference between FM ($\uparrow \uparrow$) and AFM ($\uparrow \downarrow$)
configurations of spins in the bond $ij$:
$J_{ij}=(E^{\uparrow \downarrow}_{ij}-E^{\uparrow \uparrow}_{ij})/2$S$^2$, where S=1.
Since O-$2p$ and V-$t_{2g}$ bands are separated by the large energy gap
(see figure~\ref{cavo_dos}), we consider only the interactions caused by effective transfer integrals $t^{mm'}_{ij}$
and neglect the direct contribution of the oxygen states.
In the case of CaV$_2$O$_4$, there are two electrons residing on six spin-orbitals of $t_{2g}$ symmetry.
Therefore, in the atomic limit, two majority-spin orbitals $\phi^1$ and $\phi^2$ are occupied and
all other orbitals (such as majority-spin $\phi^3$ and all minority-spin orbitals) are empty.
Then, taking into account that the hoppings are allowed only between orbitals with
the same spin, we will have:
\begin{equation}
E^{\uparrow \uparrow}_{ij}=-\frac{t_{ij}^{13}t_{ji}^{31} + t_{ij}^{23}t_{ji}^{32}}{U-3J_H} + (i\leftrightarrow j)
\end{equation}
and
\begin{equation}
E^{\uparrow \downarrow}_{ij}=-\frac{t_{ij}^{11}t_{ji}^{11}
+t_{ij}^{22}t_{ji}^{22}}{U}
-\frac{t_{ij}^{12}t_{ji}^{21}
+ t_{ij}^{21}t_{ji}^{12}
+ t_{ij}^{13}t_{ji}^{31}
+t_{ij}^{23}t_{ji}^{32} }{U-2J_H} + (i\leftrightarrow j).
\end{equation}

  Using the values of transfer integrals, collected in table III and IV of supplementary material avaliable at~\cite{SM} 
(note also that $t_{ij}^{mm'} = t_{ji}^{m'm}$)
as well as the parameters of on-site Coulomb ($U$) and exchange ($J_H$) interactions, one can obtain
that for the leg of the V1 chain: $E^{\uparrow \uparrow} =-$$3.34$ meV
and $E^{\uparrow \downarrow} =-$$42.75$ meV. Therefore, $J^{l}_1({\rm SE})$ in the superexchange approximation
can be estimated as $J^{l}_1({\rm SE}) =-$$19.7$ meV, which is in excellent agreement with $J^l_1=-$$19.9$ meV,
obtained using the theory of infinitesimal spin rotations. For the V2 chain we obtain:
$E^{\uparrow \uparrow} =-$$12.65$ meV and $E^{\uparrow \downarrow} =-$$37.89$ meV, which yield $J^{l}_2({\rm SE}) =-$$12.62$ meV,
being also in good agreement with
$J^l_2 =-$$13.86$ meV, derived from the theory of infinitesimal spin rotations. Hence the difference between
the leading exchange integrals for two nonequivalent types of vanadium reflects the behavior of transfer integrals.
The analysis for other bonds $ij$ can be performed in a similar way (details can be found in supplementary material avaliable at~\cite{SM}).
In general, we obtain a good agreement between results of the superexchange theory and the one of the infinitesimal
spin rotations.

  The experimental estimations of the exchange interactions in CaV$_2$O$_4$ have been performed in two ways.
On the one hand, the high temperature susceptibility data
have been fitted using S=1 chain model with the nearest-neighbor and next-nearest-neighbor interactions. In notations
of our paper, they corresponds to $J^{\pm}$ and $J^l$, respectively.
The solution of this model using the exact diagonalization method leads
to the $J^{\pm}({\rm SC}) = -$$19.82$ meV and $J^{l}({\rm SC})$=0~\cite{Niazi2009}, which corresponds to the linear
S=1 Haldane chains. The coupling between these chains was estimated to be $J_\perp/J^{\pm} \gtrsim  0.04$,
which corresponds to $|J_\perp({\rm SC})| \gtrsim 0.8$ meV.
Shortly after, similar fitting revealed two possible solutions with $J^{\pm}({\rm SC}) =-$$19.85$ meV, $J^{l}({\rm SC})=-$$0.75$ meV
and $J^{\pm}({\rm SC}) = -$$3.02$ meV, $J^{l}({\rm SC}) =-$$18.60$ meV~\cite{Pieper2009}.
In fact these two solutions are magnetically equivalent: in the first case $J^{\pm}({\rm SC})$ prevails and
the single spin-1 chain is realized,
while in the second case $J^{l}({\rm SC})$ is dominant, that corresponds to the formation of two independent spin-1 chains.
This illustrates the fact that the fitting of the magnetic susceptibility data for materials
with competing magnetic interactions is not unique: different sets of parameters can lead to similar behavior
of the susceptibility. The inelastic neutron scattering measurements of the magnetic excitation
spectrum in single crystals might settle this issue.

  The comprehensive analysis of complex spin wave spectrum obtained by inelastic neutron scattering (INS) technique \cite{pieper_diploma}
in low temperature monoclinic phase of CaV$_2$O$_4$ have been carried out within linear spin-wave theory and leads to the determination
of ten exchange parameters as well as two single ion anisotropy for nonequivalent V ions.
The best fit to experimental INS data was obtained for the following set of magnetic couplings:
$J^{l}_1({\rm INS}) = J^{l}_2({\rm INS}) =-$$30$ meV, $J^{+}_1({\rm INS}) =-$$11$ meV, $J^{-}_1({\rm INS}) =-$$7.9$ meV,
$J^{+}_2({\rm INS}) =7.8$ meV, $J^{-}_2({\rm INS}) =5.7$ meV~\cite{pieper_diploma}.
The full set of parameters in comparison with the one calculated in present work can be found
in table VI of supplementary material avaliable at~\cite{SM}.
These results show that the leading exchange interaction is along the leg of the zigzag chains,
that partly resolve the controversy with the fitting of the magnetic susceptibility data.
However, the value of the leg coupling $J^{l}({\rm INS}) =-$$30$ meV obtained within the spin-wave model~\cite{pieper_diploma}
is about 40\% larger than the one
derived from the fitting of the susceptibility data $J^{l}({\rm SC}) =-$$18.60$ meV~\cite{Pieper2009}.
Moreover, the exchange interactions in zigzag rungs
are rather strong $|J^{\pm}_{1,2}({\rm INS})|\approx$ 5.7-11 meV,
while the values defined by the susceptibility fitting are much smaller $|J^{\pm}({\rm SC})| =3.02$ meV.
Thus, it is clear that there is some controversy in the analysis of exchange interactions derived from the
magnetic susceptibility and inelastic neutron scattering measurements.

  To summarize this section, our theoretical value of the exchange integral $J^l_1 =-$$19.9$ meV
(for the V1 chain) is in excellent agreement with the value obtained by the fitting of susceptibility data~\cite{Niazi2009, pieper_diploma}.
Although calculated exchange parameters $|J^{\pm}_{1,2}| \approx$ 0.4-1.6 meV are somewhat
smaller than the ones estimated from the susceptibility fitting $|J^{\pm}({\rm SC})| =3.02$ meV, the general tendency
$|J^l_{1,2}| \gg |J^{\pm}_{1,2}|$ is maintained. The exchange interaction between different zigzag chains
$|J^{\pm}_{b,c}| \approx$ 1.05-2.9 meV are also consistent with the estimation
based on the susceptibility fitting $J_{\perp} \approx  1$ meV~\cite{pieper_diploma}.
Nevertheless, our theoretical calculations reveal a strong difference of exchange interactions in the legs
of two crystallographically inequivalent chains: $J^l_1 =-$$19.9$ meV and $J^l_2 =-$$13.9$ meV.
It is also worth to mention that the experimental and theoretical exchange interactions
seem to evidence against the ladder
model ($J^l$, $J^+ \gg J^-$) for the monoclinic phase of CaV$_2$O$_4$.

\subsection{Susceptibility}

\begin{figure}[htb]
\begin{center}
\includegraphics[width=8cm, angle=-90]{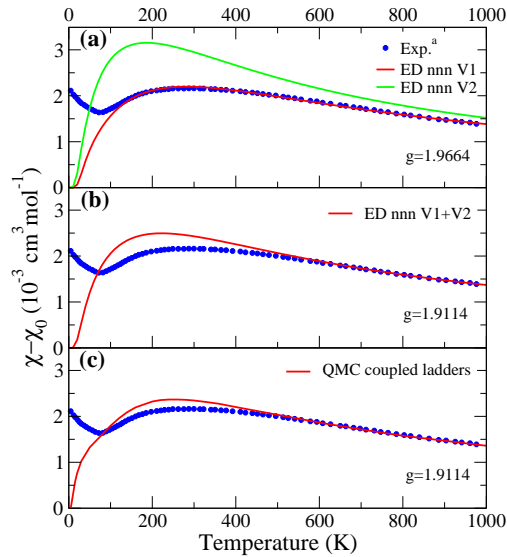}
\end{center}
\caption{\label{chi} (color online).
The comparison of experimental static magnetic susceptibility $\chi-\chi_0$ ($\chi_0$=0.48$\times$10$^{-3}$ cm$^3$/mol -- temperature independent contribution)
from~\cite{Pieper2009, pieper_diploma} (H$ \parallel c$) shown as blue dots with the
solution of Heisenberg model with calculated values of exchange interactions. Panel (a): comparison with ED solution of
next-nearest-neighbors chains of V1 and V2 (red and green curves, correspondingly). Panel (b): comparison with the sum of two
chains (red curve). Panel (c): comparison with the QMC solution of the coupled ladders model (red curve).
}
\end{figure}

  In order to compare the obtained values of the exchange interactions with experiment we first solve the next-nearest-neighbor spin-1 chain
Heisenberg model separately for V1 and V2 using exact diagonalization (ED) method implemented in the ALPS simulation package \cite{ALPS1}.
In these calculations, for the nearest-neighbor interactions in the chain $i$, we use the averaged value of $J^+_i$ and $J^-_i$;
and for the next-nearest-neighbor interactions, we use $J^l_i$. By doing so, we actually simulate the behavior of the
orthorhombic phase, which is realized above 141 K and for which $J^+_i = J^-_i$.
The L=12 spins along the chain were taken into account.
From figure~\ref{chi}(a) one can see that the behavior of the single V1 chain with the leading exchange $J^l_1 =-$$19.9$ meV agrees with experimental
data very well, in agreement with results of~\cite{Pieper2009}. Since
the $J^l_2 =-$$13.9$ meV is substantially smaller than $J^l_1$,
the susceptibility for the V2 chain is overestimated.
By considering these two noninteracting with each other zigzag chains, the total susceptibility should be obtained
by averaging the data for the individual chains. Because of the V2 contribution,
the obtained susceptibility deviates considerably from the experimental one below 500 K (see figure~\ref{chi}(b)), indicating that probably
the model of two noninteracting alternating chains is not appropriate for CaV$_2$O$_4$.

Then, we try to take into account the interactions between different chains and solve a more complex model using
quantum Monte-Carlo method implemented in the ALPS simulation
package~\cite{ALPS1,ALPS2}. Because of the complexity of the problem (the existence of two inequivalent chains and different
types of interactions between the chains) we have to rely on additional simplifications.
First, we neglect the contributions of small and alternating (FM and AFM) interactions $J^{\pm}_b$.
As the result, the problem is reduced to the analysis of a two-dimensional model. Then, we consider the ladders,
consisting of two different interactions, $J^l_1$ and $J^l_2$, in two legs of this ladder, and take into account the
strongest interchain interaction $J^-_c = 2.9$ meV as the rung of the ladder. Finally, we consider the interaction
between these ladders. For these purpose we use the average value of four parameters: $J^+_1$, $J^-_1$, $J^+_2$, and $J^-_2$.
The results of these simulations are shown in figure~\ref{chi}(c). The considered two-dimensional model substantially
improves the agreement with the experimental data and reproduces the wide peak of susceptibility at around 250 K.
The values of the $g$-factor (1.996 and 1.911), obtained from the fitting of calculated susceptibility
to the experimental data are within the typical data range $1.92  \leq g \leq  2.00$ used for the vanadium compounds and the value 1.958
obtained from Curie-Weiss fitting of the experimental susceptibility in~\cite{Pieper2009}.

\section{Conclusions}
The electronic structure, orbital configuration and magnetic interactions of quasi-one-dimensional antiferromagnet CaV$_2$O$_4$ was
studied. For these purpose, the Hubbard-type model for $t_{2g}$
states have been constructed with all the parameters derived from the first-principles calculations. The crystal field splitting and the orbital
order is found to be different for two types of 
crystallographically inequivalent vanadium atoms. This affects the behavior of interatomic exchange interactions,
which is found to be different, in several respects, from the phenomenological picture solely based on the
analysis of the crystal structure of CaV$_2$O$_4$ and fitting of the experimental magnetic susceptibility. 
Particularly, we have found that the exchange interactions in two crystallographically inequivalent zigzag chains behave rather
differently. Furthermore, there is a substantial interaction between the zigzag chains, which is comparable with intrachain
interactions. This analysis allowed us to resolve several controversial issues, regarding the leading exchange
interactions in CaV$_2$O$_4$ and the relative roles played by the intrachain and interchain interactions. 
Moreover, we argues that the interaction between the zigzag chains is an important ingredient of realistic spin model,
which should be taken into account, for instance, in the analysis of magnetic susceptibility data.

\section *{Acknowledgements}
The authors thank Prof. D. C. Johnston, B. Lake and O. Pieper for providing the comprehensive information on crystal and magnetic
structure of CaV$_2$O$_4$.
This work is supported by the project 14-12-00306 of the Russian Scientific Foundation.
The part of the calculations were performed on the ``Uran'' cluster of the IMM UB RAS.


\section*{References}
\begin{thebibliography}{10}

\bibitem{Kikuchi2001} Kikuchi H, Chiba M and Kubo T 2001 {\it Can. J. Phys.} {\bf 79} 1551

\bibitem{Niazi2009} Niazi A {\it et al.} 2009 {\it Phys. Rev. B} {\bf 79} 104432

\bibitem{Pieper2009} Pieper O {\it et al.} 2009 {\it Phys. Rev. B} {\bf 79} 180409

\bibitem{SM} Supplementary materials

\bibitem{Bertaut1967} Bertaut E F and van Nhung N and Hebd C R 1967 {\it Seances Acad. Sci. B} {\bf 264} 1416

\bibitem{Hastings1967} Hastings J M, Corliss L M, Kunnmann W and La Placa S 1967 {\it J. Phys. Chem. Solids} {\bf 28} 1089

\bibitem{Sugiyama2008} Sugiyama J, Ikedo Y, Goko T, Ansaldo E J, Brewer J H, Russo P L, Chow  K H and Sakurai H 2008 {\it Phys. Rev. B} {\bf 78} 224406

\bibitem{Zong2008} Zong X, Suh B J, Niazi A, Yan J Q, Schlagel D L, Lograsso T A and Johnston D C 2008 {\it Phys. Rev. B} {\bf 77} 014412

\bibitem{MommaK.Izumi2011} Momma K and Izumi F 2011 {\it J. Appl. Crystallogr.} {\bf 44} 1272

\bibitem{downfolding} Solovyev I V 2008 {\it J. Phys.: Condens. Matter.} {\bf 20} 293201

\bibitem{navo_prb} Pchelkina Z V, Solovyev I V and Arita R 2012 {\it Phys. Rev. B} {\bf 86} 104409

\bibitem{pieper_diploma} Pieper O  {\it Ph.D. thesis} Der Technischen Universität Berlin (2010)

\bibitem{Kugel1982} Kugel K I and Khomskii D I 1982 {\it Sov. Phys. Usp.} {\bf 25} 231

\bibitem{Kanamori1963} Kanamori J 1963 {\it Prog. Theor. Phys.} {\bf 30} 275

\bibitem{Liechtenstein} Liechtenstein A I, Katsnelson M I, Antropov V P and Gubanov V A 1987 {\it J. Magn. Magn. Matter.} {\bf 67} 65

\bibitem{Anderson1959} Anderson P W 1959 {\it Phys. Rev.} {\bf 115} 2

\bibitem{ALPS1} Bauer B {\it et al.} (ALPS collaboration) 2011 {\it J. Stat. Mech.} P05001; Albuquerque A F {\it et al.} (ALPS collaboration) 2007 
{\it Journal of Magnetism and Magnetic Materials} {\bf 310} 1187

\bibitem{ALPS2} Alet F, Wessel S and Troyer M 2005 {\it Phys. Rev. E} {\bf 71} 036706

\end{thebibliography}
\end{document}